\documentclass[preprint,amsmath,amssymb,aps]{revtex4-1}

\usepackage{graphicx}
\usepackage{dcolumn}
\usepackage{bm}
\usepackage{braket}

\usepackage{amsmath}	
\usepackage{subcaption}
\usepackage{multirow}

\begin{document}

\title{1/2$^-$ $\alpha$ cluster resonances of $^{\bf 13}{\bf C}$ studied by the analytic continuation in the coupling constant}
\author{Seungheon Shin}
 \affiliation{Department of Physics, Hokkaido University, Sapporo 060-0810, Japan}
\author{Masaaki Kimura}
 \email{masaaki.kimura@riken.jp}
  \affiliation{RIKEN Nishina Center, Wako, Saitama 351-0198, Japan}
\author{Bo Zhou}
 \affiliation{Key Laboratory of Nuclear Physics and Ion-Beam Application (MoE), Institute of Modern Physics, Fudan University, Shanghai 200433, China}
 \affiliation{Shanghai Research Center for Theoretical Nuclear Physics, NSFC and Fudan University, Shanghai 200438, China}
\author{Qing Zhao}
 \affiliation{School of Science, Huzhou University, Huzhou 313000, Zhejiang, China}

\date{\today}

\begin{abstract}
The 1/2$^-$ resonant states in $^{13}{\rm C}$ are investigated to search for the Hoyle-analog state. In order to treat the resonance states located around the 3$\alpha+n$ threshold, the analytic continuation in the coupling constant (ACCC) has been combined with the real-time evolution method (REM). The properties of the 1/2$^-$ resonance states such as the radii and monopole transition probabilities are calculated. We show the 1/2$^-_3$ and 1/2$^-_4$ states are well-developed $\alpha$ cluster states, and the 1/2$^-_4$ state is a candidate of the Hoyle-analog state.
\end{abstract}

\pacs{Valid PACS appear here}
\maketitle


\section{Introduction}
The Hoyle state, the second 0$^+$ state of $^{12}{\rm C}$, is located just above the 3$\alpha$ breakup threshold, and it is one of the most well-known and important $\alpha$ cluster states. It is considered to have a 3$\alpha$ gas-like structure, i.e., the Bose-Einstein condensate (BEC), in which the $\alpha$ particles mostly condensate in the 0$S$ orbit \cite{tohsaki2001a}. The study of the Hoyle state has provided new insights into the dynamics of $\alpha$ clusters \cite{Funaki2003,Chernykh2007,Kanada-Enyo2007,Itoh2011,Epelbaum2011,Epelbaum2012,Freer2012,Fukuoka2013,Ohtsubo2013,funaki2015hoyle,Carlson2015,Zhou2016}. In recent years, the concept of the $\alpha$-particle condensate has been extended to the 4$\alpha$ \cite{Wakasa2007,funaki2008alpha,itoh2010search,ohkubo2010alpha,rodrigues2014,funaki2015cluster}, 5$\alpha$ \cite{itagaki2008coupling,adachi2021candidates} and many $\alpha$ \cite{yamada2004dilute,kokalova2006signatures,kawabata2013search,bishop2019experimental} cluster systems. The Hoyle-analogue states in non-$N\alpha$ nuclei are also quite important issues to understand the dynamics of $\alpha$-particle BEC \cite{ogloblin2010measuring,rodrigues2014resonant,danilov2015study}.

The $^{13}{\rm C}$ nucleus is an interesting research target because it potentially has the cluster structure composed of  3$\alpha$ clusters plus one neutron \cite{kawabata2008cluster,Furutachi2011,yamada2015alpha,Chiba2020,inaba2021search,shin2021isoscalar}. The Hoyle-analog state in $^{13}{\rm C}$ is expected to have the valence neutron in the lowest $s$ or $p$ orbit coupled to the Hoyle state, which forms $J^\pi=1/2^\pm$ or 3/2$^-$ states. Recently, Kawabata $et~al.$ \cite{kawabata2008cluster} observed the strong isoscalar monopole transitions from the ground state to the excited 1/2$^-$ state around $E_x$ = 12.5 MeV. The same state was also reported by Inaba $et~al.$ \cite{inaba2021search}. Because this state is a mirror state of $^{13}{\rm N}$ at 13.5 MeV that dominantly decays to the Hoyle state \cite{fujimura2004nuclear}, it is a strong candidate of the Hoyle-analog state. In addition, Inaba $et~al.$ \cite{inaba2021search} measured strong isoscalar dipole transitions and suggested the 16.1 MeV state with $J^\pi=1/2^+$ or 3/2$^+$ as another candidate for the $\alpha$ condensed state.

On the other hand, theoretical studies have been made by Chiba $et~al.$ \cite{Chiba2020} using the antisymmetrized molecular dynamics (AMD) and by Funaki $et~al.$ \cite{yamada2015alpha} using the orthogonality condition model (OCM). They argued that the Hoyle analog state exists as the 1/2$^+$ states approximately at 15.4 and 14.9 MeV, respectively. As for the $J^\pi=1/2^-$ state, Chiba $et~al.$ \cite{Chiba2020} reported that some of the excited states showed the 3$\alpha+n$ configuration, but their spectroscopic factors for the Hoyle state channel were too small to be regarded as the Hoyle analog state, and Funaki $et~al.$ \cite{yamada2015alpha} concluded that there is no $1/2^-$ state which can be regarded as a Hoyle-analog state. Thus, theoretical studies of the Hoyle-analog state are still controversial, and further investigation is required.

In the previous work \cite{shin2021isoscalar}, we have investigated the candidates of the Hoyle-analog state with $J^\pi=1/2^-$ by using the real-time evolution method (REM) \cite{Imai2019}. REM has successfully described the various cluster states in several light nuclei \cite{Zhou2020,shin2021shape,zhao2022microscopic,zhao2022alpha}. However, due to the contamination of the continuum, we could not identify resonances at high excitation energies. Since the Hoyle-analog state is expected to be embedded in the 3$\alpha+n$ continuum, accurate treatment of the resonances is essential. 

In this study, to identify the resonant states, we combine the analytic continuation in the coupling constant (ACCC) \cite{kukulin1989,tanaka1999unbound} with REM. The aim of this work is two-fold. The first is the benchmark of combining ACCC with REM. The second is the investigation of the 1/2$^-$ resonant states of $^{13}{\rm C}$ for which the experimental and theoretical arguments are not consistent.

We organize this paper as follows. In the next section, the frameworks of REM and ACCC are outlined. In Sec.~\ref{sec:result}, we first discuss the benchmark calculations of $^{8}{\rm Be}$ and $^{5}{\rm He}$ to test the combination of the ACCC and REM. Based on this, the 1/2$^-$ resonant states of $^{13}{\rm C}$ are investigated. The calculated radii and monopole strengths within ACCC provide structural information of the 1/2$^-$ resonances. We propose that the 1/2$^-_3$ and 1/2$^-_4$ states are well-developed $\alpha$ cluster states, and especially, the 1/2$^-_4$ state is worth investigating as a candidate of the Hoyle-analog state in $^{13}{\rm C}$. In the last section, this work is summarized.

\section{Theoretical framework}
\subsection{Real-time evolution method}

We here outline the Hamiltonian and the framework of REM for $\alpha$ clusters plus a valence neutron system. The Hamiltonian for the $N\alpha+n$ nuclear system is given as,
\begin{align}
 H = \sum_{i=1}^{4N+1} t_i- t_{cm} + \sum_{i<j}^{4N+1}v_N(r_{ij}) +
 \sum_{i<j}^{4N+1}v_C(r_{ij}) , \label{eq:ham}
\end{align}
where $t_i$ and $t_{cm}$ denote the kinetic energies of the $i$th nucleon and the center-of-mass, respectively. The $v_{N}$ and $v_{C}$ are the effective nucleon-nucleon interaction and Coulomb interaction. The nucleon-nucleon interaction $v_{N}$ includes the central force and the spin-orbit interaction. In this study, we use the Minnesota force \cite{thompson1977systematic} and Reichstein and Tang spin-orbit interaction \cite{reichstein1970study} for $^{8}{\rm Be}$ and $^{5}{\rm He}$, and the Volkov No.2 \cite{Volkov1965} and G3RS \cite{Yamaguchi1979} interactions for $^{13}{\rm C}$. Especially, as for $^{13}{\rm C}$, we have uncertainty in the Hamiltonian because we cannot fix the strength of the G3RS interaction as it gives two possible choices, V$_{ls}$ = 800 and 2000 MeV. All the parameters will be explained in the next section. It is noted here that the Reichstein and Tang spin-orbit interaction has the form of,
\begin{align}
  V_{ij}=-V_\gamma e^{-\gamma r^2_{ij}}(\bm{r}_i-\bm{r}_j)\times(\bm{p}_i-\bm{p}_j)\cdot(\bm{\sigma}_i+\bm{\sigma}_j)\frac{1}{2\hbar}.
\end{align}
We will take the zero-range limit and introduce an alternative strength parameter $J_{ls}=V_\gamma\gamma^{-5/2}$ for $^{5}{\rm He}$ as in Ref.~\cite{thompson1977systematic}.

The Brink-Bloch wave function~\cite{Brink1966} is employed as the intrinsic wave function. The wave function for the $N\alpha+n$ system can be written as, 
\begin{align}
 \Phi(\bm Z_1,...,\bm Z_{N+1}) &= \mathcal A
 \Set{\Phi_\alpha(\bm Z_1)\cdots\Phi_\alpha(\bm Z_N)\phi_n(\bm Z_{N+1})},
 \label{eq:brink1}\\ 
 \Phi_\alpha(\bm Z) &= \mathcal A
 \Set{\phi(\bm r_1,\bm Z)\chi_{p\uparrow}\cdots\phi(\bm r_4,\bm Z)\chi_{n\downarrow}},\\
 \Phi_n(\bm Z) &=\phi(\bm r,\bm Z)\chi_{n\uparrow},\label{eq:brink2}\\
 \phi(\bm r,\bm Z) &= \left(\frac{2\nu}{\pi}\right)^{3/4}\exp
 \set{-\nu\left(\bm r- \bm Z\right)^2},
\end{align}
where $\Phi_\alpha(\bm Z)$ and $\Phi_n(\bm Z)$ are the wave functions of an $\alpha$ particle and a neutron located at $\bm Z$, respectively. The vectors $\bm Z_1,...,\bm Z_{N+1}$ represent the coordinates and momenta of nucleons. We fix the neutron spin of the intrinsic wave function to up, which does not affect the generality. The different Gaussian width parameter $\nu$ are used for $^{8}{\rm Be}$, $^{5}{\rm He}$ and $^{13}{\rm C}$, which will be explained in the next section.

In the REM framework, the basis wave functions are generated from the equation-of-motion (EOS), which describes various configurations of clusters. The EOS is derived from the time-dependent variational principle,  
\begin{align}
 \delta\int dt\frac{\langle\Phi(\bm Z_1,...,\bm Z_{N+1})|i\hbar\;d/dt-H|
 \Phi(\bm Z_1,...,\bm Z_{N+1})\rangle}{\langle\Phi(\bm Z_1,...,\bm Z_{N+1})|
 \Phi(\bm Z_1,...,\bm Z_{N+1})\rangle}=0. 
\end{align}
The EOM for the Gaussian centroids $\bm Z_1,...,\bm Z_{N+1}$ is obtained as,
\begin{align}
  &i\hbar\sum_{j=1}^{N+1}\sum_{\sigma=x,y,z} C_{i\rho j\sigma}\frac{dZ_{j\sigma}}{dt} =
 \frac{\partial \mathcal H_{int}}{\partial Z_{i\rho}^*}, \label{eq:eom}\\ 
 \mathcal H_{int}&\equiv\frac{\langle\Phi(\bm Z_1,...,\bm Z_{N+1})|H|
 \Phi(\bm Z_1,...,\bm Z_{N+1})\rangle}{\langle\Phi(\bm Z_1,...,\bm Z_{N+1})|
 \Phi(\bm Z_1,...,\bm Z_{N+1})\rangle},\\
 C_{i\rho j\sigma}&\equiv\frac{\partial^2\text{ln}\langle\Phi(\bm Z_1,...,\bm Z_{N+1})|
 \Phi(\bm Z_1,...,\bm Z_{N+1})\rangle}{\partial Z^*_{i\rho}\partial Z_{j\sigma}}.
\end{align}
It generates a set of the vectors $\{\bm Z_1(t),...,\bm Z_{N+1}(t)\}$ as a function of real-time $t$. We note that the EOM (Eq.~(\ref{eq:eom})) keeps the intrinsic energy $\mathcal H_{int}$, which is the energy calculated with the intrinsic wave functions (Eq.~(\ref{eq:brink1})), constant during the time evolution. In the practical calculation, we introduce an external trap field so that the constituent particles move in a finite size space \cite{shin2021shape}. This external field changes the momentum of a particle at a large distance which we call the rebound radius. It is set to 20 fm for $^8$Be and $^5$He, and 10 fm for $^{13}$C. 

A set of the basis wave functions generated by EOM are projected to the eigenstates of parity and angular momentum, and superposed. In other words, this is a generator coordinate method that employs the real-time $t$ as a generator coordinate.
\begin{align}
 \Psi^{J\pi}_M=\int_0^{T_{\text{max}}}dt \sum^J_{K=-J}\hat{P}_{MK}^{J\pi}f_K(t)
 \Phi(\bm Z_1(t),...,\bm Z_{N+1}(t)).\label{eq:gcmwf}
\end{align}
Here, the parity and the angular momentum projection operator is given as
\begin{align}
 \hat{P}^{J\pi}_{MK}=\frac{2J+1}{8\pi^2}\int
 d\Omega\mathcal{D}_{MK}^{J*}(\Omega)\hat{R}(\Omega)\frac{1+\pi \hat{P}_x}{2},\hspace{0.5cm}\pi=\pm,
\end{align}
where ${D}_{MK}^{J}$, $R(\Omega)$, and $P_x$ denote Wigner D-matrix, rotation and parity operators, respectively. In the practical calculation, the integral in Eq.~(\ref{eq:gcmwf}) is discretized as,
\begin{align}
 \Psi^{J\pi}_M = \sum_{iK}\hat{P}^{J\pi}_{MK}f_{iK}\Phi(\bm Z_1(t_i),...,\bm Z_{N+1}(t_i)). \label{eq:gcmwf2}
\end{align}
By solving the Hill-Wheeler equation \cite{Hill1953,Griffin1957}, the amplitude $f_{iK}$ and eigenenergy are determined.

\subsection{Analytic continuation in the coupling constant}

Here, we explain the ACCC \cite{kukulin1989,tanaka1999unbound}. The ACCC Hamiltonian $H'$ is comprised as,
\begin{align}
    H'(\lambda)=H+\lambda H_a,
\end{align}
where $H$ is the original Hamiltonian of the physical system and $H_a$ is an auxiliary potential multiplied by the coupling constant $\lambda$. The Hamiltonian $H'$ is identical to the physical Hamiltonian $H$ at $\lambda=0$, which we call the physical point. As $\lambda$ increases, $H'$ earns additional attractive potential. At a certain value of $\lambda$, which we denote $\lambda_0$, a resonance becomes a bound state. The trajectory of the eigen-energy as a function of $\lambda$ in the bound state region with $\lambda>\lambda_0$ is analytically continuated to the unbound region with $\lambda<\lambda_0$ to evaluate the resonance energy and width at the physical point. In the practical calculation, following the procedure made in Ref.~\cite{kukulin1989}, we fit the trajectory by a fractional function as,
\begin{align}
    \frac{\hbar^2}{2m}k^2(x)=&\,E(x),\\
    k(x):=&\,i\,\frac{p_0+p_1x+p_2x^2+\cdots+p_Nx^N}{1+q_1x+q_2x^2+\cdots+q_Mx^M},\hspace{0.5cm}\text{with }x:=\sqrt{\lambda-\lambda_0},\label{eq:pade}
\end{align}
where $m$ is the reduced mass, and $N+M+1$ coefficients $p_1,...,p_N,q_1,...,q_M$ are determined by the fitting. In the present calculations, $N$ and $M$ are chosen as $N = M = 4$ for $^8$Be, 13 for $^5$He, and 6 for $^{13}{\rm C}$. Once we fit the analytic function $k(x)$, we continuate it to the physical point. At the physical point $x=i\sqrt{\lambda_0}$, the wave number,
\begin{align}
    k(i\sqrt{\lambda_0})=k_\text{r}-ik_\text{i},\hspace{0.5cm}k_\text{r},k_\text{i}>0,
\end{align}
which gives the energy and width of a resonance,
\begin{align}
    E=\frac{\hbar^2k^2}{2m}:= E_R-i\Gamma/2,\hspace{0.5cm} E_R=\frac{\hbar^2}{2m}(k_\text{r}^2-k_\text{i}^2),\hspace{0.5cm}\Gamma=\frac{2\hbar^2}{m}k_\text{r}k_\text{i}.
\end{align}
For the case of the $s$-wave state, we need a special treatment as discussed in Ref.~\cite{kukulin1989}. 

\section{results and discussion}\label{sec:result}
\subsection{Benchmark calculations of ACCC with REM}

We first confirm the validity of combining the REM framework with ACCC. Two-body cluster systems, $^8$Be($\alpha+\alpha$) and $^5$He($\alpha+n$), are compared with the preceding ACCC results by Tanaka $et~al.$ \cite{tanaka1999unbound}. For this purpose, we employ the same Hamiltonian used in Ref.~\cite{tanaka1999unbound}. The parameter of the Minnesota potential is chosen as $u=0.94$ for $^8$Be and $u=0.98$ for $^5$He. The zero-range limit for the spin-orbit interaction is adopted, and its strength parameter $J_{ls}=50$ MeV$\cdot$fm$^2$ was applied. The size parameter of the Gaussian wave packet $\nu=0.26$ fm$^{-2}$ is used.
To solve EOM, we choose the intrinsic energies $\mathcal H_\text{int}=13.8$ and 18 MeV for $^8$Be and $^5$He, respectively, so that the wave functions can cover large model space which is essential for the convergence of the GCM calculations. The numbers of the basis wave functions are 75 and 90 for the $^8$Be and $^5$He, respectively.  

We first show the numerical results obtained by using the Minnesota parameter $u$ as the coupling constant $\lambda$, which is the same procedure employed by Tanaka $et~al.$~\cite{tanaka1999unbound}. It is noted that increasing $u$ does not affect the energy of the $\alpha$ particle. The results are listed in Table.~\ref{accc_bench}. The results for the 0$^+$ and 2$^+$ states of $^8$Be are consistent with those reported by Tanaka $et~al.$ \cite{tanaka1999unbound}. We additionally tested the 4$^+$ state and found that the result is comparable to the experimental value in both the energy and width. As for the $^5$He case, the spin-orbit interaction yields 3/2$^-$ and 1/2$^-$ resonances which are also well reproduced in our framework. The $s$-wave state of $^5$He, which has a very large width, is also properly described and consistent with the result obtained by Tanaka $et~al.$ \cite{tanaka1999unbound} although the Pad$\text{\'{e}}$ approximation of the $s$-wave state is unstable.

\begin{table}[tbh!]
\caption{The resonance energy $E_R$ and the decay width $\Gamma$ of $^8$Be and $^5$He in the unit of MeV obtained by the ACCC + REM compared with the previous study \cite{tanaka1999unbound}. Numbers in the parenthesis are the result obtained by using the spin-orbit interaction as an auxiliary potential instead of the central interaction. Experimental data are taken from Ref.~\cite{TILLEY2004155,TILLEY20023} for $^8$Be and $^5$He, respectively.} \label{accc_bench}
\begin{ruledtabular}
\begin{tabular}{llllllll}
                     &      & \multicolumn{2}{c}{REM} & \multicolumn{2}{c}{Tanaka $et~al.$ \cite{tanaka1999unbound}} & \multicolumn{2}{c}{EXP}\\
                     & $J^\pi$ & $E_R$          & $\Gamma$      & $E_R$               & $\Gamma$ & $E_R$               & $\Gamma$           \\
                     \hline
$^8$Be & 0$^+$   & 0.224      & 0.001      & 0.208           & 0.003    &  0.09184  & 5.57 $\pm$ 0.25 eV \\
                     & 2$^+$   & 2.87       & 1.42       & 2.85            & 1.44    &  3.1218   & 1.513 $\pm$ 0.015 \\
                     & 4$^+$   & 11.77      & 4.82       &                 &       &  11.44  & $\approx$ 3.5   \\
                     \hline
$^5$He & 3/2$^-$ & 0.78(0.73)       & 0.66(0.64)       & 0.77            & 0.64    &  0.735  & 0.648 \\
                     & 1/2$^-$ & 1.98(1.93)       & 5.62(5.46)       & 1.98            & 5.4      & 2.005 & 5.57 \\
                     & 1/2$^+$ & 12.7       & 163        & 12              & 180         &   &
\end{tabular}
\end{ruledtabular}
\end{table}



\begin{figure}[hbt!]
  \begin{subfigure}{0.49\textwidth}
    \centering
    \includegraphics[width=1.05\hsize]{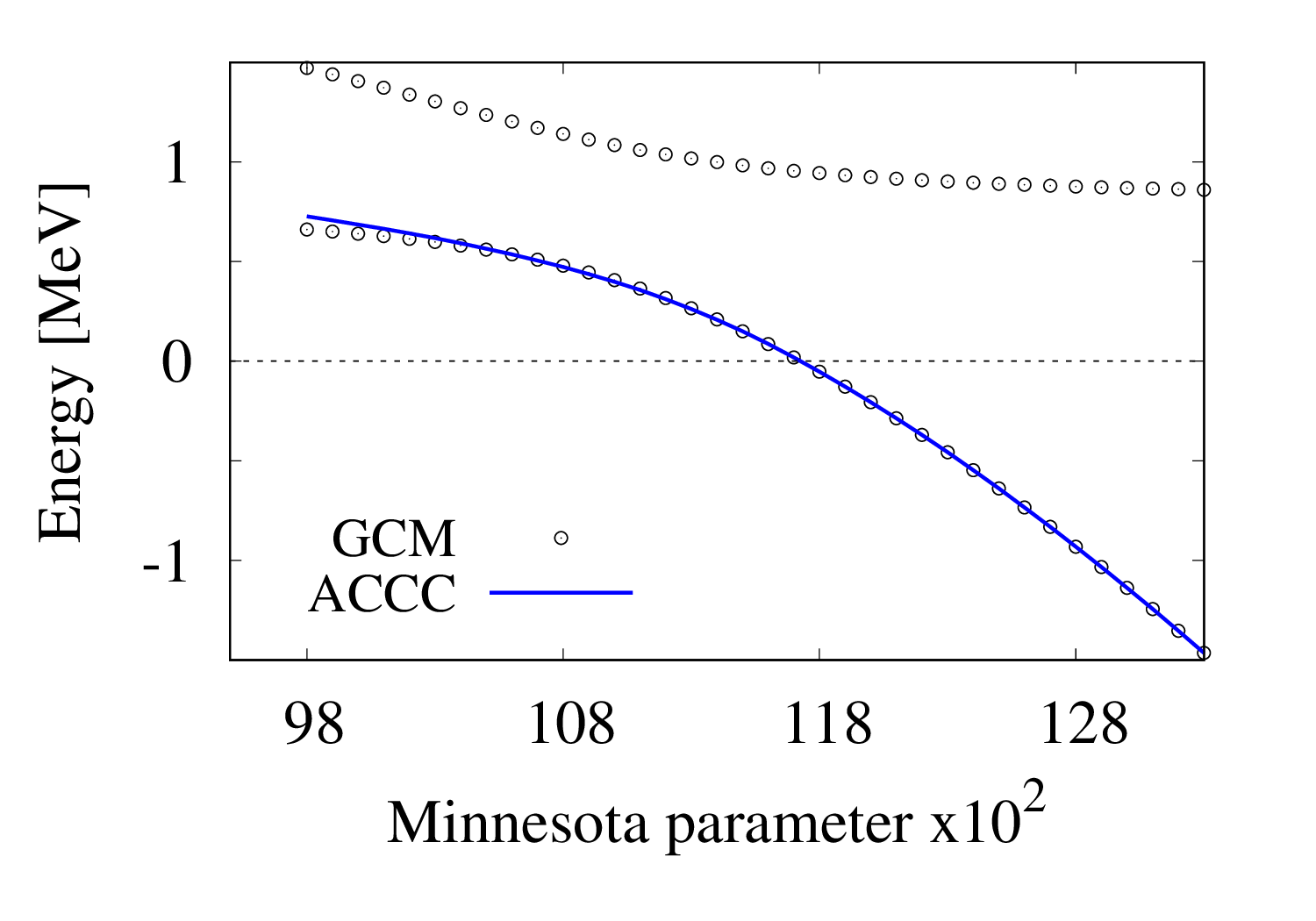}
    \caption{$\lambda=u$}
    \label{fig:lam_u}
  \end{subfigure}
  \begin{subfigure}{0.49\textwidth}
    \centering
    \includegraphics[width=1.05\hsize]{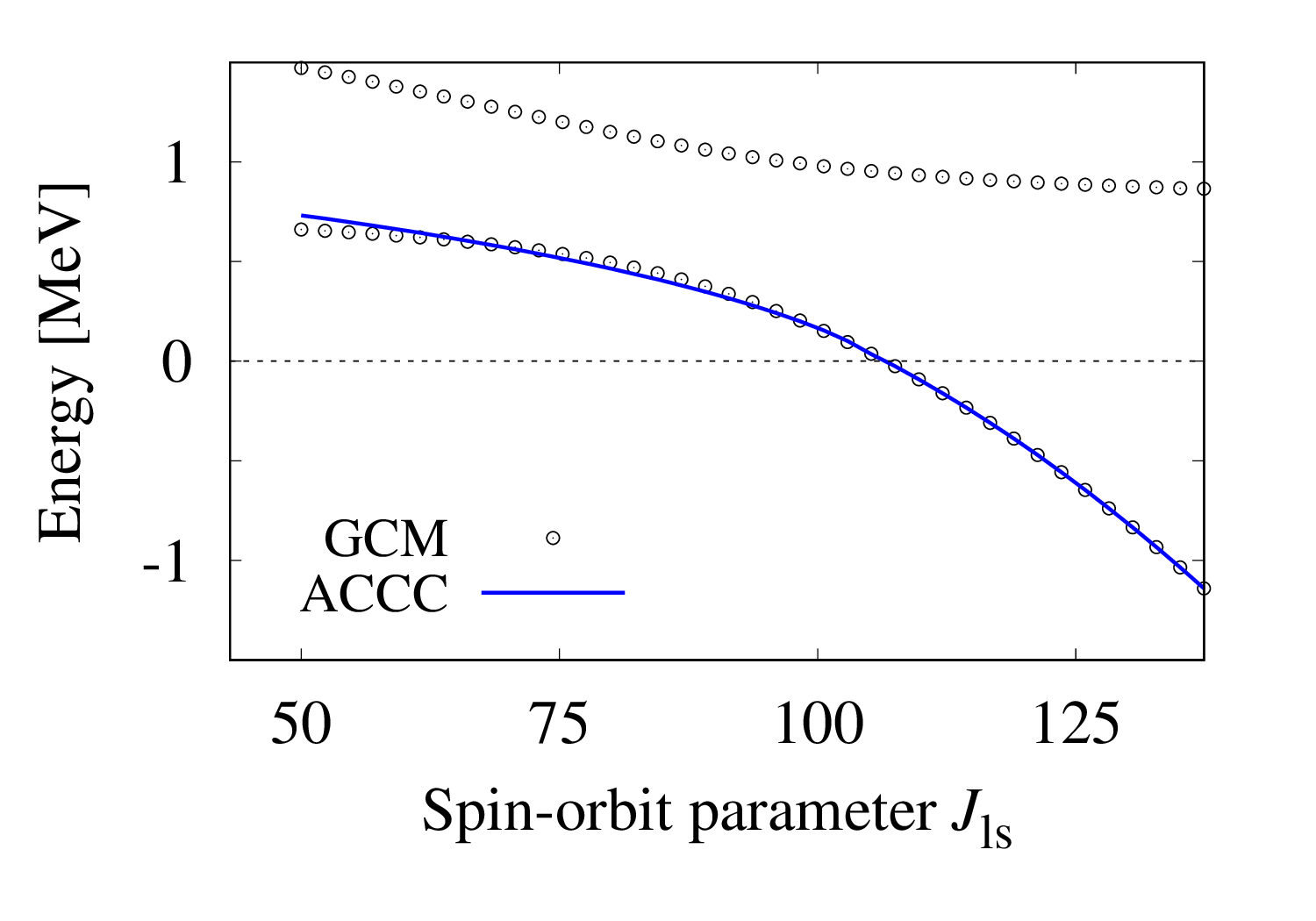}
    \caption{$\lambda=J_{ls}$}
    \label{fig:lam_J}
  \end{subfigure}
\caption{The 3/2$^-$ state of $^5$He obtained by (a) the central force (Minnesota parameter $u$ as coupling constant) and (b) the spin-orbit interaction (spin-orbit parameter $J_{ls}$ as coupling constant) as $H_a$, respectively.}
\label{fig:lam_compar}
\end{figure}

\begin{figure}[hbt!]
  \begin{subfigure}{0.49\textwidth}
    \centering
    \includegraphics[width=1.05\hsize]{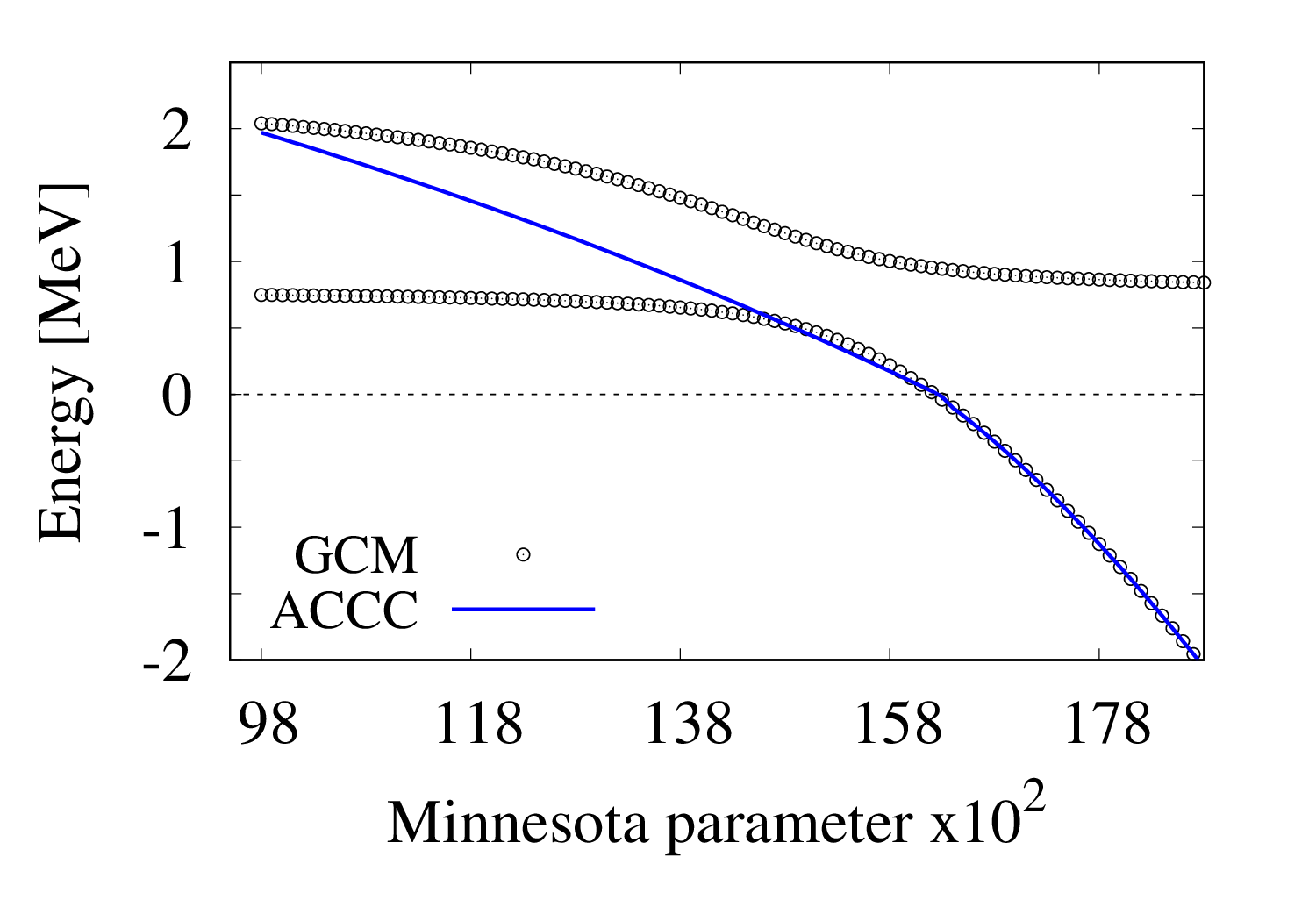}
    \caption{$\lambda=u$}
  \end{subfigure}
  \begin{subfigure}{0.49\textwidth}
    \centering
    \includegraphics[width=1.05\hsize]{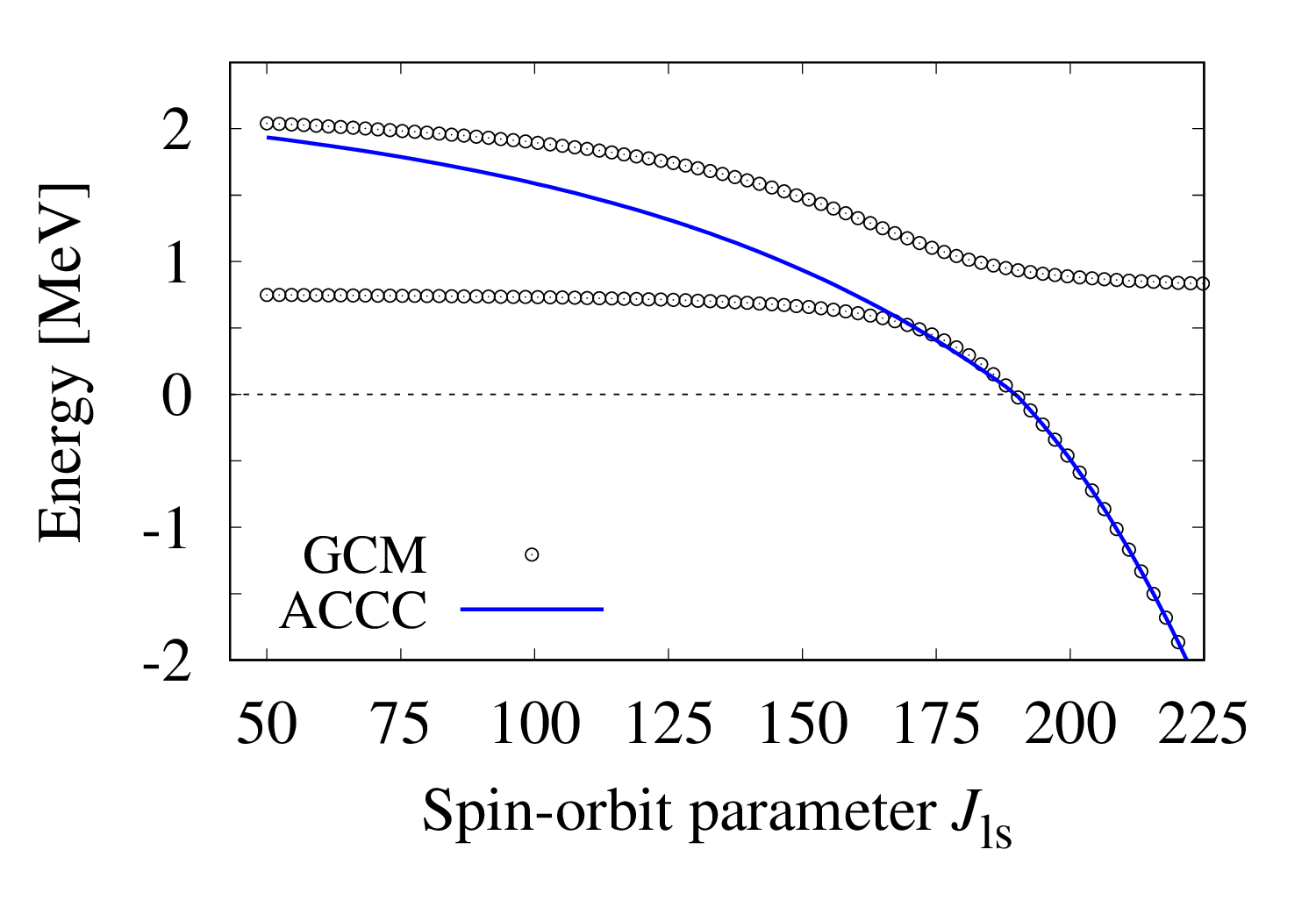}
    \caption{$\lambda=J_{ls}$}
  \end{subfigure}
\caption{Same as Fig.~\ref{fig:lam_compar}, but for the 1/2$^-$ state.}
\label{fig:lam_compar2}
\end{figure}

Since the choice of the auxiliary potential $H_a$ is arbitrary, we have also tested the spin-orbit interaction as $H_a$, and used its strength $J_{ls}$ as the coupling constant $\lambda$. It was applied to the 1/2$^-$ and 3/2$^-$ resonances of $^5$He, and the numerical results are listed in the parenthesis of Tab.~\ref{accc_bench}. The graphical comparison of the eigenvalue trajectories between the central and spin-orbit interactions as $H_a$ is shown in Fig.~\ref{fig:lam_compar} and Fig.~\ref{fig:lam_compar2}. Despite the different choices of the auxiliary potential, we could obtain reasonable results in both cases. We note that this may be the first example that uses the spin-orbit interaction as $H_a$, and we expect that this choice of the auxiliary potential will make it easy to investigate the $N\alpha+xn$ systems.



\subsection{$\bm{1/2^-}$ resonances of $^{13}$C}

In this study, we focus on the 1/2$^-$ states of $^{13}$C since the structures of the excited 1/2$^-$ states still have no consensus in both theoretical and experimental aspects. $^{13}$C is described as the 4-body system of $3\alpha+n$ in the REM framework. The resonance energy is calculated with respect to the ground state of $^{12}$C as $k=\sqrt{E(^{13}\text{C})-E(^{12}\text{C}_{\text{g.s.}})}$. When we choose the central force as $H_a$, we need to use a large value of $\lambda$ to make the resonance states of $^{13}$C bound, which makes it difficult to extrapolate to $\lambda=0$ by the Pad$\text{\'{e}}$ series. Therefore, as we proposed and demonstrated above, we use the spin-orbit interaction as $H_a$. In this case, the energy of $^{12}$C is independent of $\lambda$, which stabilizes the ACCC calculations. 

The Volkov No.2 \cite{Volkov1965} and G3RS \cite{Yamaguchi1979} interactions are used for $^{13}$C as in our previous study \cite{shin2021shape}. The Majorana parameter $M$ is set to 0.592, which gives reasonable excitation energy of the Hoyle state as 7.68 MeV (experimentally 7.65 MeV). The $\alpha$ width parameter is fixed to $\nu=0.23$ fm$^{-2}$, which reproduces the observed $\alpha$ particle's size. As for the real-time evolution calculations, two intrinsic energies are used, $\mathcal H_{int}=30$ and $40$ MeV, to cover the various configurations of the $\alpha$ particles and a neutron. We prepared 300 wave functions of each intrinsic energy, and 600 bases in total are superposed.

\begin{figure}[b!]
    \centering
    \includegraphics[width=0.9\hsize]{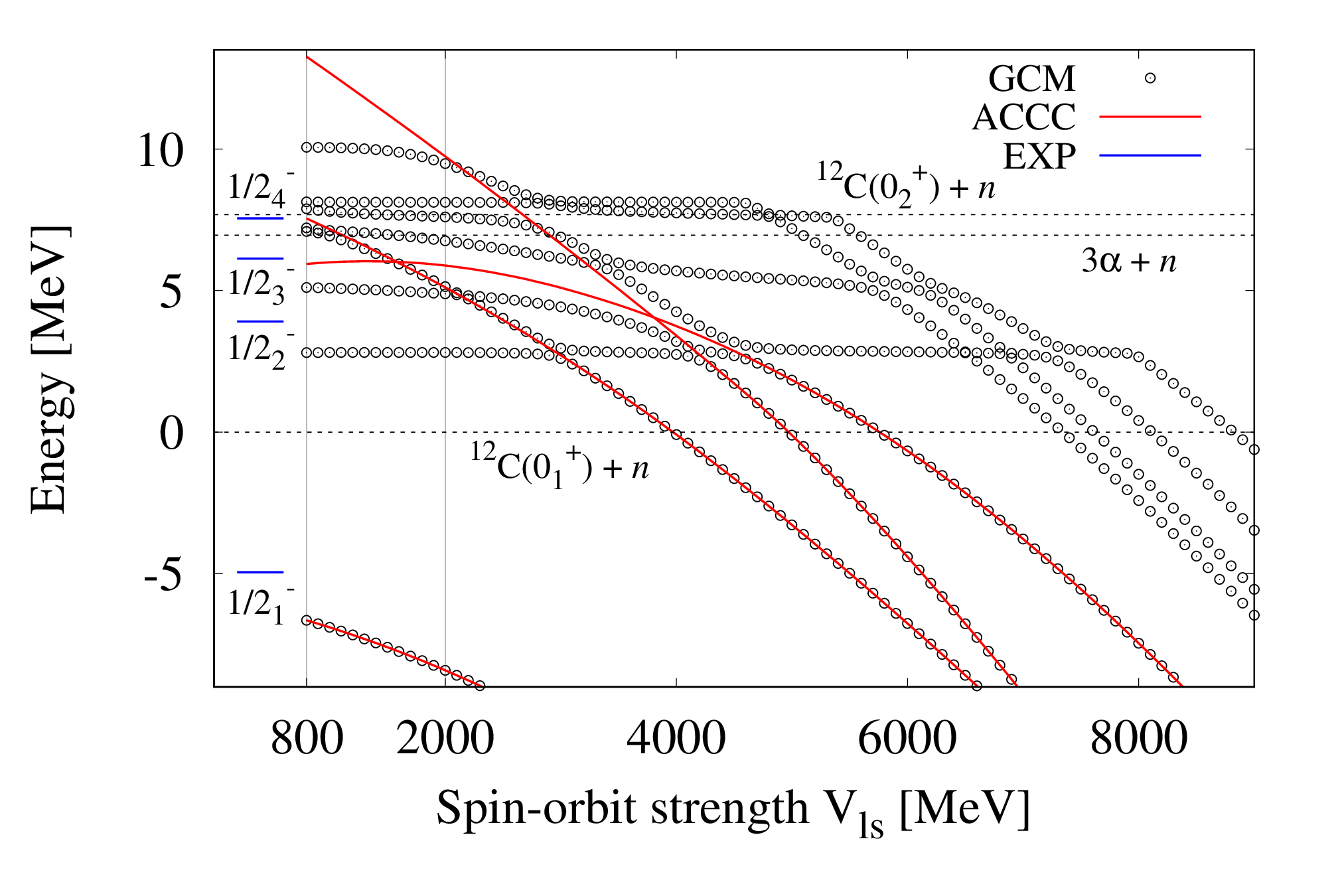}
    \caption{The 1/2$^-$ resonance states obtained by the Pad$\text{\'{e}}$ approximation.}
    \label{fig:13C_accc}
\end{figure}

Fig.~\ref{fig:13C_accc} shows the energies of the 1/2$^-$ states obtained by GCM and ACCC. Here, we have 2 choices of the physical point of the spin-orbit strength, V$_{ls}$ = 800 and 2000 MeV. The former value is the original strength of the G3RS interaction \cite{tamagaki1968potential} and the latter is chosen to reproduce the phase shift of the $\alpha+n$ system \cite{okabe1979structure}.
The numerical results are summarized in Table \ref{accc_sum} with the resonance energy $E$ and width $\Gamma$. In addition to the energy and width we also calculate the matrix elements of isoscalar $M(IS0)$ and electric $M(E0)$ monopole transition strength that can be obtained within the ACCC \cite{kukulin1989},
\begin{align}
    (\Phi|\hat{O}|\psi\rangle=\underset{\lambda\rightarrow0}{\text{Cont}}\int^\infty_0 \Phi^*(k(\lambda),r) O\psi(r)dr,\label{eq:accc_mat}
\end{align}
where the round bra with $\Phi$ denotes the resonance state and the angled ket with $\psi$ denotes the ground state. The sign `Cont' means the continuation of the coupling constant $\lambda$ to the physical point. The matter $R_m$ and proton $R_p$ radius of a resonance state can be calculated by the general form of Eq.~(\ref{eq:accc_mat}),
\begin{align}
    (\Phi_i|\hat{O}|\Phi_i)=\underset{\lambda\rightarrow0}{\text{Cont}}\int^\infty_0 \Phi^*_i(k_i(\lambda),r) O\Phi_i(k_i(\lambda),r)dr,
\end{align}
The normalization of resonance states can be obtained by Zel'dovich regularization \cite{kukulin1989}.

\begin{table}[tbh!]
\caption{Numerical results of the 1/2$^-$ states are listed. Energies $E$ from the threshold, widths $\Gamma$ of resonance states, electric ($E$0) and isoscalar ($IS$0) monopole transition densities, and matter $R_m$ and proton $R_p$ radii with the choice of V$_{ls}$ = 800 and 2000 MeV. Energy and width are in MeV, monopole transitions are in ($e$)fm$^2$, and radii are in fm$^2$. Experimental data are taken from Ref.~\cite{Ajzenberg-Selove1991}}\label{accc_sum}
\begin{ruledtabular}
\begin{tabular}{llllllll|lll}
      & \multicolumn{7}{c|}{ACCC}  & \multicolumn{3}{c}{EXP}           \\
$J^\pi$    &  V$_{ls}$   & $E$     & $\Gamma$    & M(IS0) & M(E0) & $R_m$ & $R_p$ & $E$     & $\Gamma$        & M(IS0) \\
\multirow{2}{*}{1/2$^-_1$} &800& -6.64  & -     & -      & -     & 2.43 & 2.37 & \multirow{2}{*}{-4.95} & \multirow{2}{*}{-}                & \multirow{2}{*}{-}      \\
                           &2000& -8.41  & -     & -      & -     & 2.40 & 2.36 & & &\\ \hline
\multirow{2}{*}{1/2$^-_2$} &800& -      & -     & -      & -     & -    & -    & \multirow{2}{*}{3.91}  & \multirow{2}{*}{0.15}             & \multirow{2}{*}{6.1}    \\
                           &2000& 5.11   & 0.24 & 7.11 & 3.55 & 2.80 & 2.73 & & &\\  \hline
\multirow{2}{*}{1/2$^-_3$} &800& 5.94   & 1.38 & 15.41 & 5.04 & 3.02 & 2.90 & \multirow{2}{*}{6.13}  & \multirow{2}{*}{\textless{}0.004} & \multirow{2}{*}{4.2}    \\
                           &2000& 5.88   & 0.65 & 13.52 & 3.83 & 2.93 & 2.65 & & &\\  \hline
\multirow{2}{*}{1/2$^-_4$} &800& 7.55   & 0.65 & 7.55 & 3.55 & 2.88 & 2.78 & \multirow{2}{*}{7.55}  & \multirow{2}{*}{-}                & \multirow{2}{*}{4.9}    \\
                           &2000& 9.74   & 4.80 & 6.46 & 2.85 & 2.88 & 2.88 & & &\\  \hline
\multirow{2}{*}{1/2$^-_5$} &800& 12.69 & 7.65 & 5.30  & 2.43 & 3.00 & 2.90 & \multirow{2}{*}{-} & \multirow{2}{*}{-} & \multirow{2}{*}{-} \\  
                           &2000& -     & -    & -     & -    & -    & -    & & &\\
\end{tabular}
\end{ruledtabular}
\end{table}

%
%

The spin-orbit strength V$_{ls}$ = 800 MeV reasonably reproduces the ground state. Although the 1/2$^-_2$ state is not described with V$_{ls}$ = 800 MeV, the 1/2$^-_3$ and 1/2$^-_4$ resonance energies are obtained in good agreement with the experiment. As for the 1/2$^-_5$ state, its energy is too high in this case, so we do not discuss it in detail here.

The isoscalar monopole transitions from the ground state to the excited 1/2$^-_3$ and 1/2$^-_4$ states are strong in accordance with the experiments. At the same time, the electric monopole transitions also show large values, and especially the 1/2$^-_4$ state has about half of each $IS$0 strength, which shows the spatial excitation of protons is almost the same as that of neutrons that means having clear alpha cluster structure. It is also seen that the proton radius is comparable with the matter radius.

The other choice V$_{ls}$ = 2000 MeV can reproduce the experimental energies of all the 1/2$^-$ states. The $IS$0 and $E$0 transitions again show large values where the 1/2$^-_2$ and 1/2$^-_4$ states show about the half values of $E$0 strengths of $IS$0 transitions implying the large proton distribution comparable to the neutrons. However, the proton radius is almost the same as the matter radius in the 1/2$^-_4$ state while the proton radius is much smaller in the 1/2$^-_2$ state. An interesting point is that the matter and proton radii of the 1/2$^-_4$ state are almost independent of the choice of V$_{ls}$.

Between these two possible choices of V$_{ls}$, we consider that V$_{ls}$ = 800 MeV gives a more natural description for $^{13}$C including the ground state and the other states although the 1/2$^-_2$ excited state is missing. It is also noted that a similar spin-orbit strength, V$_{ls}$ = 1000 MeV, gave a reasonable description of the low-lying band structure in our previous work \cite{shin2021shape}. 
The stronger strength, V$_{ls}$ = 2000 MeV, made each state reveal its dynamics of spatial structure. In the 1/2$^-_3$ and 1/2$^-_4$ states, the role of the valence neutron seems different, where the spin-orbit interaction is significant among the clusters in the 1/2$^-_3$ state, which shrinks the proton radius from 2.90 fm to 2.65 fm, while it almost does not affect the proton radius in the 1/2$^-_4$ state. The physical properties of the 1/2$^-_4$ state such as monopole transitions and radii are not dependent on the spin-orbit strength that implies the 1/2$^-_4$ state is a different type from the usual cluster states. We speculate that this 1/2$^-_4$ state, located above the 3$\alpha+n$ threshold, could be a candidate for the Hoyle-analog state.

The recent experimental study by Inaba $et~al.$ \cite{inaba2021search} found that there is a bump structure, which is composed of several states located close to each other, around 12.5 MeV from the ground state (the 1/2$^-_4$ exp.~state), and this bump structure, especially its $IS$0 transitions, was not described by the shell model calculation. Furthermore, this state is regarded as a mirror state of the bump structure at 13.5 MeV of $^{13}{\rm N}$ whose decay to the Hoyle state was observed \cite{fujimura2004nuclear}.
We regard that our 1/2$^-_4$ state corresponds to this experimental finding and thus, it needs to have further discussions to examine if the 1/2$^-_4$ state can be the Hoyle-analog state.

\section{summary}
In summary, the resonance states of the 1/2$^-$ states in $^{13}{\rm C}$ have been investigated. As a benchmark calculation to demonstrate the combination of ACCC and REM with the employed interactions, $^8{\rm Be}$ and $^5{\rm He}$ resonance states were well reproduced compared to the preceding ACCC study. In addition, a tactical treatment of ACCC was introduced, in which the spin-orbit interaction was exploited as an auxiliary potential in the ACCC and it properly reproduced the $^5{\rm He}$ resonance states. 

Based on this new treatment, the resonance states of $^{13}{\rm C}$ were clearly figured out from the continuum. The matrix elements of radius and monopole transitions were also calculated within the ACCC, which provides cluster characteristics of each resonance state. The physical interaction was chosen with the spin-orbit strength V$_{ls}$ = 800. The 1/2$^-_2$ state was not obtained in this study implying that this state is probably not an $\alpha$ cluster state. The 1/2$^-_3$ and 1/2$^-_4$ states were found to be the well-developed $\alpha$ cluster states, and especially the 1/2$^-_4$ state was independent of the choices of V$_{ls}$ = 800 and 2000 MeV, which exhibits the characteristic of the condensate state, so further investigation on the 1/2$^-_4$ state is required as the future work.

\begin{acknowledgements}
 This work was supported by JST SPRING, Grant Number JPMJSP2119.
\end{acknowledgements}

\bibliography{references.bib}
\end{document}